\documentclass[aps,prd,reprint,superscriptaddress]{revtex4-1}
\usepackage{amsmath}
\usepackage{amssymb}
\usepackage{bm}
\usepackage{mathrsfs}
\usepackage[pdftex,colorlinks,hyperindex,plainpages=false,bookmarksopen,bookmarksnumbered,pdfusetitle]{hyperref}

\newcommand{\he}[1]{{#1}^\dagger}
\newcommand\symd[1]{\overset\leftrightarrow{#1}}
\newcommand\DD{\mathscr{D}}
\newcommand\dd{\mathrm{d}}
\newcommand\g{\mathfrak{g}}
\newcommand\G{\mathcal G}
\renewcommand\H{\mathcal H}
\newcommand\h{\mathfrak{h}}
\newcommand\imag{\mathrm{i}}
\newcommand\A{\mathcal{A}}
\newcommand\La{\mathscr{L}}
\newcommand\vek[1]{\bm{#1}}

\begin{document}
\title{General coordinate invariance in quantum many-body systems}

\author{Tom\'{a}\v{s} Brauner}
\email{brauner@hep.itp.tuwien.ac.at}
\affiliation{Institute for Theoretical Physics, Vienna University of Technology, Vienna, Austria}
\affiliation{Department of Theoretical Physics, Nuclear Physics Institute of the ASCR, \v Re\v z, Czech Republic}

\author{Solomon Endlich}
\affiliation{Institut de Th\'eorie des Ph\'enom\`enes Physiques, EPFL, Lausanne, Switzerland}

\author{Alexander Monin}
\affiliation{Institut de Th\'eorie des Ph\'enom\`enes Physiques, EPFL, Lausanne, Switzerland}

\author{Riccardo Penco}
\affiliation{Department of Physics and ISCAP, Columbia University, New York, USA}

\begin{abstract}
We extend the notion of general coordinate invariance to many-body, not necessarily relativistic, systems. As an application, we investigate nonrelativistic general covariance in Galilei-invariant systems. The peculiar transformation rules for the background metric and gauge fields, first introduced by Son and Wingate in 2005 and refined in subsequent works, follow naturally from our framework. Our approach makes it clear that Galilei or Poincar\'e symmetry is by no means a necessary prerequisite for making the theory invariant under coordinate diffeomorphisms. General covariance merely expresses the freedom to choose spacetime coordinates at will, whereas the true, physical symmetries of the system can be separately implemented as ``internal'' symmetries within the vielbein formalism. A systematic way to implement such symmetries is provided by the coset construction. We illustrate this point by applying our formalism to nonrelativistic $s$-wave superfluids.
\end{abstract}

\maketitle


\section{Introduction}
\label{sec:intro}

Symmetries play a key role in modern understanding of the fundamental laws of nature. While they paved the way to the discovery of the standard model of elementary particles as well as to a geometric understanding of gravity, they have also proven extremely helpful as a tool for practical computations. In quantum field theory, symmetries of a system can be conveniently encoded in terms of invariance of the generating functional under transformations of a set of background fields, or external perturbations. This approach is particularly fruitful in combination with effective field theory (EFT) techniques, allowing one to connect descriptions of the same physics at vastly different length or momentum scales, possibly based of completely different dynamical degrees of freedom. A prime example of the use of generating functional methods within EFT is the chiral perturbation theory of quantum chromodynamics~\cite{Gasser:1983yg,*Gasser:1984gg}.

Conserved currents associated with physical symmetries can be probed by introducing a set of external gauge fields. Provided gravity is not involved and only \emph{internal} symmetries are of interest, it is straightforward to make the classical action invariant under simultaneous local transformations of the gauge and matter (that is, all other, non-gauge) fields by a suitable choice of transformation rules~\footnote{We tacitly assume that no obstructions to gauging are present, such as quantum anomalies or central charges in the Lie algebra of the symmetry group.}. This in turn leads to gauge invariance of the generating functional under transformations of the background fields alone~\cite{Leutwyler:1993gf,*Leutwyler:1993iq}. Spacetime symmetries, if desired, can be described by their action on field components in a fixed coordinate grid.

Once gravity enters the game, or when background fields associated with \emph{spacetime} symmetries are needed, the implementation of symmetries in the generating functional becomes nontrivial, as is evidenced by the recent discussion of general coordinate invariance in nonrelativistic (NR) systems~\cite{Son:2005rv,Son:2008ye,Son:2013rqa,Andreev:2013qsa,Abanov:2014ula,*Gromov:2014gta,Banerjee:2014pya,*Banerjee:2014nja,Cho:2014vfl,Geracie:2014nka,Gromov:2014vla,*Bradlyn:2014wla}. Here, it helps to keep in mind the viewpoint of differential geometry. Namely, treating spacetime as a (differentiable) manifold, physical observables should be viewed as geometric quantities, independent of the choice of local spacetime coordinates. Likewise, the basic building blocks of the action of a given theory are geometric objects such as scalar or vector fields, which can be defined in a coordinate-free manner. From this viewpoint, it is rather clear that general coordinate invariance merely encodes the freedom to choose a coordinate system at will. As physics must be independent of such a choice, true, physical symmetries of the system act directly upon the fields, without a reference to the spacetime coordinates. The vielbein, or frame field, formalism is particularly suited for making this distinction clear: once expressed in terms of the vielbein basis, all observables can be cast in a manifestly coordinate-free fashion. 

The objective of this paper is to clarify some of the issues pertinent to general coordinate invariance in NR systems. We do so by insisting on the above differential-geometric picture. A systematic use of the vielbein formalism enables us to keep covariance under coordinate diffeomorphisms manifest at all stages. The noncovariant transformation rules for the background fields~\cite{Son:2005rv}, including the sources for energy density and current~\cite{Son:2008ye}, emerge naturally without the need to start from a relativistic theory and perform a NR reduction. We demonstrate in particular that making a given theory generally coordinate invariant does not require global Poincar\'e or Galilei symmetry. This is because in our approach, general coordinate invariance is \emph{not} a physical symmetry to start with, and it only determines how the auxiliary background fields enter the action~\cite{Janiszewski:2012nb}.

This basic scheme determines the plan of the paper. In Sec.~\ref{sec:GCI}, we build a generally covariant framework applicable to a large class of local quantum field theories. As a nontrivial illustration, this is applied in Sec.~\ref{sec:NRGCI} to the problem considered in Refs.~\cite{Son:2005rv,Son:2008ye}, that is, a NR Galilei-invariant field theory with a U(1) internal symmetry. At this stage it could appear that Galilei invariance is imposed by a naive extension of the free Schr\"odinger field case. The rest of the paper is devoted to its systematic implementation, based on the coset construction~\cite{Coleman:1969sm,*Callan:1969sn} and its generalization to spacetime symmetries~\cite{Volkov:1973vd,*Ogievetsky}. First, in Sec.~\ref{sec:galilei} we provide a basic overview of the structure of the Galilei group and discuss how it can be implemented within the coset formalism. The next two sections then work out two particular realizations of Galilei symmetry, suitable for the description of a microscopic theory of a charged matter field (Sec.~\ref{sec:microscopic}), and of an EFT for its low-temperature, superfluid phase (Sec.~\ref{sec:superfluid}). Finally, in Sec.~\ref{sec:conclusions} we conclude and provide an outlook on possible future application of our formalism.


\subsection{Relation to recent literature}
\label{subsec:relation}

Since several papers have recently appeared which address the same or a similar problem, it is mandatory to clarify the relation of our results to these works. In Ref.~\cite{Son:2013rqa}, a framework based on the Newton--Cartan geometry was put forward as a means to construct actions manifestly invariant under \emph{spatial} diffeomorphisms. Our approach lends a simple interpretation to the structure introduced therein, both from a mathematical (in terms of the vielbein) and a physical (in terms of sources for momentum density and current) viewpoint.

Formally, the material of Sec.~\ref{sec:GCI} is related to that of Ref.~\cite{Andreev:2013qsa} which, nevertheless, did not interpret the background fields in terms of sources for energy and momentum currents. Our argument in Sec.~\ref{sec:GCI} is close in spirit to Ref.~\cite{Banerjee:2014pya,*Banerjee:2014nja} which, too, showed how to obtain a generally coordinate invariant action from one invariant under global symmetries alone. However, while we systematically use the covariant vielbein resulting in a clear and concise picture,  a similar result was obtained in~\cite{Banerjee:2014pya,*Banerjee:2014nja} by a combination of an educated guess and ``long algebra''.

While this paper was in preparation, Refs.~\cite{Geracie:2014nka,Gromov:2014vla,Bradlyn:2014wla} appeared which partially overlap with the material presented here, in particular, by illuminating the noncovariant transformation rules for some of the background fields based on the underlying geometry~\cite{Geracie:2014nka}, and by systematically using the vielbein formalism~\cite{Gromov:2014vla,Bradlyn:2014wla}. However, we believe that we provide deeper insight in some important points such as: (i) the manifestly covariant definition of the U(1) gauge field; (ii) the importance of spacetime symmetries for the relations between  conserved currents; (iii) the systematic implementation of Galilei symmetry using the coset formalism.


\subsection{Index conventions}

Throughout the paper, we use four types of indices: $\mu,\nu,\dotsc$ for spacetime coordinates, $i,j,\dotsc$ for spatial coordinates, $A,B,\dotsc$ for internal (vielbein) spacetime coordinates, and $a,b,\dotsc$ for internal spatial coordinates. Where necessary, the temporal coordinate (both internal and spacetime) is denoted by the index $0$. 


\section{General coordinate invariance}
\label{sec:GCI}

Consider a theory of a set of matter fields, denoted collectively as $\psi$, and possibly a set of background fields $A_\mu$, coupled to conserved currents of some internal symmetries of the theory. We assume that in flat space, the action of the theory, $S\{\psi,A\}$, is invariant under simultaneous gauge transformations of the matter and gauge fields. Provided there is no anomaly introduced by functional integration, this in turn gives rise to a gauge-invariant generating functional,
\begin{equation}
Z\{A\}\equiv\int\mathcal D\psi\,e^{\imag S\{\psi,A\}}.
\end{equation}
We now wish to couple the theory to background spacetime geometry. One possible motivation for doing so may be that the space(time) actually is curved. More often, the physical picture that we will have in mind is rather a system living in flat spacetime. Here then, the spacetime fluctuations serve merely as source fields for conserved currents of spacetime symmetries, analogous to $A_\mu$, which allow for a concise description of correlators of these currents.


\subsection{Vielbein formalism}
\label{subsec:vielbein}

We are relatively free in the choice of the background fields: different choices may result in generating functionals with different sets of symmetries, which give us access to different observables. It will prove convenient to work with the spacetime vielbein, $e^A_\mu(x)$, representing a fixed local basis in which spacetime tensors can be decomposed. Under an infinitesimal coordinate diffeomorphism, $x'^\mu=x^\mu+\xi^\mu(x)$, the vielbein transforms as a one-form, that is
\begin{equation}
\Delta e^A_\mu=-e^A_\nu\partial_\mu\xi^\nu.
\label{vielbeintransfo}
\end{equation}
For the sake of brevity, we used here a shorthand notation for the form variation of a field, $\Delta\psi(x)\equiv\psi'(x')-\psi(x)$. The total local variation is thus given by $\delta\psi(x)\equiv\psi'(x)-\psi(x)=\Delta\psi(x)-\xi^\mu(x)\partial_\mu\psi(x)$. In flat spacetime, we can always choose the local basis as $e^A_\mu\to\delta^A_\mu$, hence we will often be interested in the vielbein fluctuation, $A^A_\mu\equiv\delta^A_\mu-e^A_\mu$. By Eq.~\eqref{vielbeintransfo}, this satisfies the transformation rule
\begin{equation}
\Delta A^A_\mu=\partial_\mu\xi^A-A^A_\nu\partial_\mu\xi^\nu.
\label{sourcetransfo}
\end{equation}
This is exactly what we would expect from a one-form gauge field for local translations induced by coordinate diffeomorphisms. The setup is completed by introducing the dual vielbein, $E^\mu_A$, defined as
\begin{equation}
E^\mu_A e^B_\mu=\delta^B_A.
\label{dualvielbein}
\end{equation}
Under diffeomorphisms, this naturally transforms as a vector, $\Delta E^\mu_A=E^\nu_A\partial_\nu\xi^\mu$.

The covariant vielbein allows us to make any action, defined by a local Lagrangian, invariant under coordinate diffeomorphisms. The only assumption we make is that the Lagrangian is expressed solely in terms of field variables that can be given a geometric meaning without referring to a particular coordinate frame. This means in other words that the Lagrangian is composed out of scalar, vector and tensor fields and otherwise does not depend on the coordinates explicitly. The construction of the generally covariant Lagrangian out of its counterpart defined in flat spacetime then proceeds in three steps:
\begin{enumerate}
\item\textit{Scalar fields.} It is convenient to first turn all field variables into spacetime scalars, which is done by projecting their spacetime tensor components onto the vielbein or its dual. For instance, in the case of a vector field $\tilde{\psi}^\mu$ this amounts to introducing the scalar components $\psi^A\equiv e^A_\mu\tilde\psi^\mu$.
\item\textit{Covariant derivatives.} Suppose that for any scalar field $\psi$ (internal indices suppressed), we can construct a covariant derivative $D_\mu\psi$ that transforms under all internal symmetries like $\psi$ itself. Notice that, whenever $\psi$ itself has a vielbein index attached to it (either because it carries spin, or because it is a covariant derivative of another field) this may not be an easy task. Defining the covariant derivative $D_\mu\psi$ in general is done most simply using the coset approach introduced later in this paper. If, however, $\psi$ is a single spacetime scalar, the curved background does not introduce any subtlety. Then, using the dual vielbein, the covariant derivative $D_\mu \psi$ can be projected to a diffeomorphism-scalar derivative,
\begin{equation}
\DD_A\psi\equiv E^\mu_AD_\mu\psi.
\label{scalarderivative}
\end{equation}
\item\textit{Volume measure.} By applying the above steps iteratively, we can convert any local Lagrangian density into a scalar under diffeomorphisms. Then, one can obtain  an invariant action by introducing an appropriate volume measure, defined as usual by the determinant of the vielbein: $\dd t\,\dd\vek x\,\|e_\mu^A\|$.
\end{enumerate}
This procedure should be thought of as the analogy of minimal coupling to gauge fields of internal symmetries. There is no \emph{unique} way to introduce gauge fields into a given theory with global symmetry. As long as the spacetime geometry is treated merely as a background source, this is not a problem, however: \emph{any} action invariant under diffeomorphisms, which reduces to the same action in a trivial background, or flat spacetime, is equally good.

Hopefully, it is now clear that, contrary to the impression one might get from the literature, making the theory diffeomorphism-invariant is the easy part of the problem. The truly nontrivial step in the construction of the action is the implementation of true, \emph{physical} symmetries, which provides a prescription of how to contract the vielbein indices $A$. In the following sections of this paper we will elucidate exactly how to systematically encode the physical symmetries inherent in a given system.


\subsection{Nonrelativistic notation}

Despite using the Lorentz-like spacetime indices $\mu$ and $A$, we have not made any assumptions about the actual spacetime symmetries of the system, apart from translation invariance in flat spacetime which is implicit in the requirement that the Lagrangian does not depend explicitly on the coordinates. However, since we have in mind primarily applications to NR many-body systems, it may be more transparent to treat spatial and temporal indices separately. Thus, we represent the vielbein as $e^A_\mu=(n_\mu,e^a_\mu)$ and its dual by $E^\mu_A=(V^\mu,E^\mu_a)$. The one-form $n_\mu$ can be thought of as defining surfaces of constant time, whereas $V^i$ represents a ``velocity'' variable~\cite{Son:2013rqa,Hoyos:2013eha}. The inversion relation~\eqref{dualvielbein} is now cast equivalently as
\begin{equation}
\begin{gathered}
V^\mu n_\mu=1,\qquad
V^\mu e^a_\mu=0,\\
E^\mu_a n_\mu=0,\qquad
E^\mu_a e^b_\mu=\delta^b_a.
\end{gathered}
\label{dualvielbeinNR}
\end{equation}
Splitting the spatial and temporal indices effectively casts the vielbein in a block-matrix form so that its dual, being a matrix inverse, can be computed using the block-matrix algebra~\cite{Powell:2011nq}. We thus obtain
\begin{equation}
\begin{gathered}
V^\mu=\frac{1}{n_0}(\delta^{\mu0}-E^\mu_a e^a_0),\qquad
E^0_a=-\frac{E^i_an_i}{n_0},\\
E^i_a\tilde e^b_i=\delta^b_a,\qquad\text{where}\qquad
\tilde e^a_i\equiv e^a_i-\frac{e^a_0n_i}{n_0}.
\end{gathered}
\end{equation}
The expressions for $V^\mu$ and $E^\mu_a$ become explicit once we find the inverse of the reduced spatial vielbein, $\tilde e^a_i$. Using the algebraic identity for the determinant of a block matrix, $\det\begin{pmatrix}A & B\\C & D\end{pmatrix}=\det A\det(D-CA^{-1}B)$, the volume measure can also be evaluated as
\begin{equation}
\dd x\,\|e^A_\mu\|=\dd x\,n_0\|\tilde e^a_i\|.
\label{measure}
\end{equation}


\subsection{Background field transformations}
\label{subsec:backtransfo}

Following Eq.~\eqref{sourcetransfo}, we argued that the fluctuations  $A^A_\mu$ of the vielbein  behave as gauge fields for local translations, that is, they act as sources for the energy--momentum density and current. However, since to obtain a vertex function of a conserved current, one takes a derivative with respect to the source and subsequently sets it  to zero, the precise parametrization of $e^A_\mu$ is to a large extent arbitrary. We only require that to linear order in the sources, $n_\mu\simeq\delta_{\mu0}-\delta_{\mu0}\Phi-\delta_\mu^iB_i$, denoting the sources for energy density and current as $\Phi$ and $B_i$, respectively. Here we are temporarily adopting the notation of Ref.~\cite{Son:2008ye} for easy direct comparison. One practically convenient parametrization is, for instance,
\begin{equation}
n_\mu=e^{-\Phi}(1,-B_i).
\label{ndef}
\end{equation}
From the general covariant transformation rule for the vielbein~\eqref{vielbeintransfo}, one then obtains by a straightforward manipulation the transformation of the sources,
\begin{equation}
\begin{split}
\Delta\Phi&=\partial_0\xi^0-B_k\partial_0\xi^k,\\
\Delta B_i&=\partial_i\xi^0-B_k\partial_i\xi^k+B_i(\partial_0\xi^0-B_k\partial_0\xi^k),
\end{split}
\label{Btransfo}
\end{equation}
in accord with Eq.~(38) of Ref.~\cite{Son:2008ye}. While the particular form of these transformation rules is specific to the chosen parametrization~\eqref{ndef} and thus inessential, what is important is the way to obtain manifestly covariant expressions by using the vielbein systematically.


\subsection{Comparison with Newton--Cartan formalism}

In applications, vielbeins often appear in pairs. Assuming spatial rotational invariance, we expect the covariant derivatives $\DD_a\psi$ to enter the action through expressions such as $\delta^{ab}\DD_a\he\psi\DD_b\psi=\delta^{ab}E^\mu_a E^\nu_b D_\mu\he\psi D_\nu\psi$, so it is natural to introduce a degenerate symmetric tensor
\begin{equation}
g^{\mu\nu}\equiv\delta^{ab}E_a^\mu E_b^\nu.
\label{gmunu}
\end{equation}
The spatial component of this tensor, $g^{ij}=\delta^{ab}E_a^iE_b^j$, can be inverted, giving rise to a spatial metric,
\begin{equation}
g_{ij}=\delta_{ab}\tilde e^a_i\tilde e^b_j.
\label{gijdef}
\end{equation}
Using Eqs.~\eqref{vielbeintransfo} and~\eqref{Btransfo}, and noting that in the parametri\-zation~\eqref{ndef} we have $\tilde e^a_i=e^a_i+e^a_0B_i$, one infers the transformation rule for $g_{ij}$ under diffeomorphisms,
\begin{equation}
\Delta g_{ij}=-g_{kj}\partial_i\xi^k-g_{ik}\partial_j\xi^k-(B_ig_{kj}+B_jg_{ik})\partial_0\xi^k,
\label{Deltag}
\end{equation}
again in accord with Ref.~\cite{Son:2008ye}.

Three important remarks are in order here. First, in the general case of spacetime diffeomorphisms, using the metric $g_{ij}$ creates more problems than it solves due to its complicated transformation properties. However, the situation dramatically simplifies if one restricts to (possibly time-dependent) spatial diffeomorphisms so that $\xi^0=0$, as was originally done in Ref.~\cite{Son:2005rv}. One can then set consistently $\Phi=B_i=0$, upon which the metric $g_{ij}$ transforms properly as a two-form. Spatial indices of other tensors can then be raised and lowered with $g^{ij}$ and $g_{ij}$ as usual.

Second, it is tempting to introduce another degenerate symmetric tensor which, like $g^{\mu\nu}$, does transform covariantly under all spacetime diffeomorphisms,
\begin{equation}
h_{\mu\nu}\equiv\delta_{ab}e^a_\mu e^b_\nu.
\end{equation}
We use a different symbol to emphasize that this is \emph{not} an inverse of $g^{\mu\nu}$ (which does not have an inverse to begin with). From their definition, it is easy to see that $h_{\mu\nu}$ and $g^{\mu\nu}$ satisfy a number of relations such as
\begin{equation}
g^{\mu\nu}n_\nu=0,\quad
h_{\mu\nu}V^\nu=0,\quad
g^{\mu\lambda}h_{\lambda\nu}=\delta^\mu_\nu-V^\mu n_\nu,
\label{NCconstraints}
\end{equation}
and, together with $V^\mu$ and $n_\mu$, reproduce the Newton--Cartan structure of Ref.~\cite{Son:2013rqa}. Using the relation   $V^i=-E^i_ae^a_0e^\Phi$ we see that the field $V^\mu$ introduced \emph{ad hoc} in Refs.~\cite{Son:2013rqa,Geracie:2014nka} has a simple interpretation: it is associated with the source $e^a_0$ for momentum density.

Third, we believe that the vielbein formalism offers conceptual as well as practical advantages compared to the Newton--Cartan geometry. The vielbein is obviously a fundamental object, directly related to sources for conserved currents, whereas the tensors $g^{\mu\nu}$ and $h_{\mu\nu}$ are derived. In addition, when correlators of the currents are desired, the generating functional $Z\{A_\mu,n_\mu,V^\mu,g^{\mu\nu}\}$ must be varied subject to constraints of the type~\eqref{NCconstraints}~\cite{Geracie:2014nka}. On the contrary, all components of $e^A_\mu$ are independent, hence the generating functional expressed as $Z\{A_\mu,e^A_\mu\}$ does not suffer from such problems. Most importantly, however, the Newton--Cartan formalism implicitly assumes spatial rotational symmetry in that it contracts vielbein indices using the Kronecker $\delta_{ab}$. Should we deal with an intrinsically anisotropic system, the use of the vielbein in combination with the treatment of physical symmetries, discussed later in this paper, is mandatory.


\section{Example: a Galilei-invariant system}
\label{sec:NRGCI}

In the previous section, we showed how to couple a given theory to fluctuations of spacetime geometry. We now wish to illustrate the general formalism on a concrete example as well as to investigate the constraints that physical symmetries can impose. We therefore consider a class of NR systems, described in flat spacetime by the Lagrangian
\begin{equation}
\La=\frac\imag2\he\psi\symd D_0\psi-\frac{\delta^{ij}}{2m}D_i\he\psi D_j\psi-V(\he\psi\psi).
\label{LagSchr}
\end{equation}
Here, $\psi$ is a complex scalar field that interacts with a background U(1) gauge field $A_\mu$ via minimal coupling, $D_\mu\psi\equiv(\partial_\mu-\imag A_\mu)\psi$. We also used the shorthand notation $\he\psi\symd D_\mu\psi\equiv\he\psi D_\mu\psi-D_\mu\he\psi\psi$.  The model (\ref{LagSchr}) has recently been used to describe the fractional quantum Hall effect~\cite{Geracie:2014nka}. Apart from global spacetime symmetries (spacetime translations, rotations and Galilei boosts), it is invariant under the local U(1) transformation
\begin{equation}
\Delta\psi=\imag\theta\psi,\qquad
\Delta A_\mu=\partial_\mu\theta.
\label{Atransfo}
\end{equation}

In order to cast the Lagrangian in a form independent of the choice of the coordinate frame, we simply replace gauge-covariant derivatives with the diffeomorphism-covariant ones,
\begin{equation}
\La=\frac\imag2\he\psi\symd\DD_0\psi-\frac{\delta^{ab}}{2m}\DD_a\he\psi\DD_b\psi-V(\he\psi\psi).
\label{LagImproved}
\end{equation}
We could stop at this point for we have accomplished our goal: to make the theory~\eqref{LagSchr} invariant under coordinate diffeomorphisms. However, it is instructive to make the dependence of the Lagrangian on the individual fields more explicit, as it will help us highlight the difference between diffeomorphisms and actual symmetries.

To this end, we rewrite the Lagrangian~\eqref{LagImproved} in the form put forward in Ref.~\cite{Son:2008ye},
\begin{equation}
\begin{split}
\La={}&\frac\imag2e^\Phi\he\psi\symd{\tilde D}_0\psi-V(\he\psi\psi)\\
&-\frac{g^{ij}}{2m}(\tilde D_i\he\psi+B_i\tilde D_0\he\psi)(\tilde D_j\psi+B_j\tilde D_0\psi),
\end{split}
\label{SonLagB}
\end{equation}
where we used the definition~\eqref{gmunu}. The twisted covariant derivative, $\tilde D_\mu\psi\equiv(\partial_\mu-\imag\tilde A_\mu)\psi$, is defined in terms of the appropriately modified electromagnetic field
\begin{equation}
\begin{split}
\tilde A_0&\equiv A_0+\tfrac 12mV^2n_0,\\
\tilde A_i&\equiv A_i-mV_i+\tfrac 12mV^2n_i,
\end{split}
\label{A0AiB}
\end{equation}
with the shorthand notation $V_i\equiv g_{ij}V^j$ and $V^2\equiv V^iV_i$. Notice that the \emph{form} of the field redefinition~\eqref{A0AiB} is independent of $\Phi$ and $B_i$, and hence of the chosen parametrization of $n_\mu$.

Up to differences in notation, Eq.~\eqref{A0AiB} agrees with similar relations obtained in Refs.~\cite{Hoyos:2013eha,Son:2013rqa,Geracie:2014nka}. In contrast to these papers, we \emph{began} from the covariant field $A_\mu$. For the sake of completeness, we write down the transformation rules for $\tilde A_\mu$, which are obtained by a straightforward calculation using the properties of its ingredients,
\begin{equation}
\begin{split}
\Delta\tilde A_0&=\partial_0\theta-\tilde A_\mu\partial_0\xi^\mu,\\
\Delta\tilde A_i&=\partial_i\theta-\tilde A_\mu\partial_i\xi^\mu-me^\Phi g_{ik}\partial_0\xi^k.
\end{split}
\label{Atildetransfo}
\end{equation}
These transformation rules are not a product of intuition or guesswork, but rather descend directly from our fully covariant formalism. 

The appearance of the combination $\tilde A_\mu$ in the action has a very important consequence. In general, taking a functional derivative of the action with respect to a background gauge field gives the corresponding conserved current. Now, $A_\mu$ acts as a source for the particle number current $j^\mu$, whereas $e^a_\mu$ is a source for the momentum current $T^\mu_{\phantom\mu a}$. Noting that to linear order in the sources, $A_i$ and $e^a_0$ only enter the action~\eqref{SonLagB} through the combination $A_i-mV_i$, immediately leads to the relation
\begin{equation}
T^{0i}=mj^i,
\label{T0imji}
\end{equation}
valid in the trivial background. In the absence of a source for momentum density, this relation can only be derived by a more-or-less explicit computation~\cite{Son:2005rv}. On fairly general grounds, it can be traced to the underlying Galilei symmetry~\cite{Greiter:1989qb,Son:2005rv,Brauner:2014aha}.

The above observation has a very general validity. The appearance of sources only through certain combinations is one of the hallmarks of spacetime symmetries. Whereas here we had to rely on a specific type of Lagrangian, in Sec.~\ref{sec:microscopic} we will see such combinations of sources emerge directly as a consequence of the symmetry. This leads to relations among the corresponding Noether currents, similar to Eq.~\eqref{T0imji}. The generality of such relations was already pointed out in Ref.~\cite{Brauner:2014aha}.

As should by now be clear from our construction, each of the terms in Eq.~\eqref{LagImproved} is \emph{separately} a scalar under coordinate diffeomorphisms. This might seem puzzling since we know that it is Galilei invariance what forces the temporal and spatial derivatives to appear in the particular combination featured in Eq.~\eqref{LagImproved}. However, in the vielbein formalism, the physical Galilei symmetry should \emph{not} be thought of as a finite-dimensional subgroup of the diffeomorphism group. Indeed, a much larger class of local field theories can obviously be made diffeomorphism-invariant. Diffeomorphisms merely represent the freedom of a coordinate choice, and any physical symmetry must be imposed on top of general covariance.

In this spirit, Galilei invariance of the theory~\eqref{LagImproved} can be viewed as an ``internal'' symmetry acting on the vielbein indices. Under an infinitesimal \emph{local} Galilei boost with the velocity parameter $u^a$, the covariant derivatives therein are shifted by
\begin{equation}
\DD_0\psi\to\DD_0\psi-u^a\DD_a\psi,\qquad
\DD_a\psi\to\DD_a\psi+\imag mu_a\psi.
\label{galileiguess}
\end{equation}
The former rule is identical to a boost transformation of temporal gradients, whereas the latter reflects the fact that on complex fields, Galilei boosts are realized projectively. In addition, the Lagrangian~\eqref{LagImproved} is obviously preserved by local internal rotations of the vielbein $e^a_\mu$. While here we found the symmetries by a mere educated guess, in Sec.~\ref{sec:microscopic} we will see how they can be imposed systematically using the coset construction.


\section{Spacetime symmetries}
\label{sec:galilei}

As repeatedly emphasized above, true physical symmetries can, and should, be treated separately from mere coordinate diffeomorphisms. It is a purpose of this section to set up a systematic procedure that will allow us to do so. We will follow closely the treatment of Ref.~\cite{Ivanov:1981wn} and its recent application to the coupling of spontaneously broken symmetries to gravity~\cite{Delacretaz:2014oxa}. 

The same physics can be equivalently described using the spacetime coordinates $x^\mu$ or using coordinates $y^A(x)$ defined in the local frame $e^A_\mu(x)$. In flat spacetime, one can, for instance, choose the vielbein as a global orthonormal basis, and define $y^A$ geometrically as its integral curves, $\dd y^A=e^A_\mu\dd x^\mu$. The action of symmetry transformations on the system can be described in such a basis in a way independent of the coordinate chart $x^\mu$. From the point of view of $x^\mu$, even spacetime symmetries therefore appear to be ``internal''. Any given symmetry can now be made local using standard techniques by introducing a set of background gauge fields, one for each symmetry generator. Since spacetime symmetries, in particular translations, act on the coordinates as affine rather than linear transformations though, one has to exercise some care. The implementation of the symmetries can be worked out economically using the method of nonlinear realizations, also known as the coset construction~\cite{Coleman:1969sm,*Callan:1969sn,Volkov:1973vd}. For the reader's convenience, the elements of the coset construction are reviewed in Appendix~\ref{app:coset}.


\subsection{Poincar\'e invariance}
\label{subsec:poincare}

In order to set the stage, we review here briefly the case of relativistic systems, invariant under the Poincar\'e group~\cite{Ivanov:1981wn}. We wish to describe its action in the space of local coordinates $y^A$, \emph{scalar} under diffeomorphisms. While Lorentz transformations, generated by the operator of angular momentum $J_{AB}$, act on them linearly, spacetime translations, generated by the momentum operator $P_A$, act on $y^A$ by a shift. Geometrically, one can view the Minkowski space spanned by $y^A$ as the coset space $\G/\H$, where $\G$ is the Poincar\'e group and $\H$ its Lorentz subgroup. From the general expression~\eqref{cosetparamgeneral}, the coset space can be represented by the matrix field
\begin{equation}
U(x)\equiv e^{\imag y^A(x)P_A}.
\label{cosetparam}
\end{equation}
The action of a group element $\g\in\G$ is then defined as in Eq.~\eqref{cosettransfo2}. This immediately reproduces the expected behavior under Lorentz transformations, $y^A\xrightarrow{}\Lambda^A_{\phantom AB}y^B$ [in which case $\h(y,\g)=\g$], as well as under spacetime translations, $y^A\to y^A+a^A$ [in which case $\h(y,\g)=\openone$].

The virtue of the coset construction is that it admits a straightforward gauging of the symmetry. To this end, one introduces a set of one-form gauge fields $R^{AB}_\mu$ for the generators $J_{AB}$, and $S^A_\mu$ for $P_A$. Put together in the matrix field, $\A_\mu\equiv\tfrac12 R^{AB}_\mu J_{AB}+S^A_\mu P_A$, they transform according to Eq.~\eqref{gaugetransfogeneral}. The last element of the construction is the Maurer--Cartan (MC) form~\eqref{MCformgeneral}. It can be evaluated explicitly using the commutation relations of the Poincar\'e algebra. Projecting it back to the two subspaces of generators, one finds that $\omega_\mu^{J,AB}=-R^{AB}_\mu$, whereas~\cite{Ivanov:1981wn}
\begin{equation}
\omega_\mu^{P,A}=\partial_\mu y^A-S^A_\mu-R^{AB}_\mu y_B.
\label{omegaPA}
\end{equation}
The latter quantity is invariant under translations generated by $P_A$. It transforms as a vector under the Lorentz group $\H$ and can be interpreted as the covariant vielbein, that is, $\omega_\mu^{P,A}\equiv e^A_\mu$. The connection $\omega^{J,AB}_\mu$, on the other hand, allows one to define a covariant derivative of matter fields $\psi$ that transform nontrivially under the Lorentz group $\H$ via Eq.~\eqref{MCformcovder}, i.e.
\begin{equation}
D_\mu\psi\equiv[\partial_\mu-\tfrac\imag2R_\mu^{AB}\mathcal{R}(J_{AB})]\psi.
\label{covderpoinc}
\end{equation}

The above construction leads to a generating functional $Z\{e_\mu^A ,R_\mu^{AB}\}$, invariant separately under internal gauge spacetime translations and Lorentz transformations, as well as under coordinate diffeomorphisms. The coordinates $y^A$ can be fixed arbitrarily and do not represent dynamical degrees of freedom, hence they also appear in the generating functional. Nevertheless, they only enter together with $S^A_\mu$ and $R^{AB}_\mu$ through the combination defined by Eq.~\eqref{omegaPA}. At this stage, $R^{AB}_\mu$ is still an independent background field. If desired, it can be eliminated in a way that respects all the symmetries, for instance, by setting to zero the field-strength tensor associated with $\omega^P_\mu$, that is, the torsion tensor,
\begin{equation}
\omega^{P,A}_{\mu\nu}=D_\mu\omega^{P,A}_\nu-D_\nu\omega^{P,A}_\mu.
\end{equation}
This determines $R^{AB}_\mu$ in terms of derivatives of $e^A_\mu$, giving it a value typical for a spin connection~\cite{Ivanov:1981wn},
\begin{equation}
\begin{split}
R^{AB}_\mu=-\tfrac12\bigl[&e^{\nu A}(\partial_\mu e^B_\nu-\partial_\nu e^B_\mu)+e_{\mu C}e^{\nu A}e^{\lambda B}\partial_\lambda e^C_\nu\\
&-(A\leftrightarrow B)\bigr].
\end{split}
\label{spinconnection}
\end{equation}
After this reduction, the generating functional depends solely on $e^A_\mu$ and is still constrained separately by the internal translations and spacetime diffeomorphisms. This reproduces the algorithm for construction of invariant actions outlined in Sec.~\ref{sec:GCI}. Additionally, however, the coset construction tells us that covariant derivatives for fields with spin must be defined according to Eq.~(\ref{covderpoinc}).

Notice that the internal coordinates $y^A$ were introduced merely as a device that allows us to treat spacetime symmetries separately from coordinate diffeomorphisms. At the end of the day, they can be fixed at will; a natural choice in flat spacetime is $y^A=\delta^A_\mu x^\mu$. Upon fixing the internal coordinates, geometric objects such as vectors or tensors become ``pinned'' to the coordinate grid; spacetime symmetries and coordinate diffeomorphisms then become linked to each other. However, all general arguments can be carried out without such gauge-fixing.


\subsection{Galilei invariance}
\label{subsec:galilei}

We now proceed to our main object of interest in this section: the Galilei symmetry. The generators of the Galilei algebra will be denoted as $J_{ab}$ (rotations), $N_a$ (boosts), $P_a$ (space translations), and $H$ (time translations). In the case of a single quantum-mechanical particle, the commutator $[P_a,N_b]$ has a central charge, causing the Galilei group to be realized projectively~\cite{Weinberg:1995v1}. In a many-body system composed of particles of the same mass $m$, the Lie algebra can be closed by introducing the operator of particle number $Q$. All nontrivial commutators of the Galilei algebra then read
\begin{equation}
\begin{split}
[J_{ab},J_{cd}]&=\imag(\delta_{ac}J_{bd}+\delta_{bd}J_{ac}-\delta_{ad}J_{bc}-\delta_{bc}J_{ad}),\\
[J_{ab},P_c]&=\imag(\delta_{ac}P_b-\delta_{bc}P_a),\\
[J_{ab},N_c]&=\imag(\delta_{ac}N_b-\delta_{bc}N_a),\\
[P_a,N_b]&=\imag mQ\delta_{ab},\\
[H,N_a]&=\imag P_a.
\end{split}
\label{commutators}
\end{equation}
One can associate with the above generators the gauge fields $R^{ab}_\mu$ (rotations), $B^a_\mu$ (boosts), $S^a_\mu$ (spatial translations), $T_\mu$ (time translations), and $A_\mu$ (particle number). They can all be collected in the matrix gauge field
\begin{equation}
\A_\mu\equiv\tfrac12R^{ab}_\mu J_{ab}+B^a_\mu N_a+S^a_\mu P_a+T_\mu H+A_\mu Q.
\end{equation}
The transformation of the gauge fields under local symmetry transformations is again given by Eq.~\eqref{gaugetransfogeneral}. Choosing a parametrization of the Galilei group in terms of the transformation parameters $\alpha^{ab}$, $u^a$, $a^a$, $b$ and $\theta$,
\begin{equation}
\g=\exp(\tfrac\imag2\alpha^{ab}J_{ab}+\imag u^aN_a+\imag a^aP_a+\imag bH+\imag\theta Q),
\label{gelement}
\end{equation}
we obtain explicit expressions for infinitesimal field transformations~\footnote{Whereas in Sec.~\ref{sec:GCI} the symbol $\Delta$ represented the form change of fields under coordinate diffeomorphisms, here we abuse the notation somewhat by using it for the combined action of diffeomorphisms and internal symmetries. We believe that no confusion can arise though.},
\begin{align}
\notag
\Delta R^{ab}_\mu&=\partial_\mu\alpha^{ab}-R^{ab}_\nu\partial_\mu\xi^\nu+\alpha^{a}_{\phantom ac}R^{cb}_{\mu}+\alpha^{b}_{\phantom bc}R^{ac}_{\mu},\\
\notag
\Delta B^a_\mu&=\partial_\mu u^a-B^a_\nu\partial_\mu\xi^\nu+\alpha^a_{\phantom ab}B^b_\mu-u_bR^{ab}_\mu,\\
\notag
\Delta S^a_\mu&=\partial_\mu a^a-S^a_\nu\partial_\mu\xi^\nu+\alpha^a_{\phantom ab}S^b_\mu-a_bR^{ab}_\mu+u^a T_\mu-bB^a_\mu,\\
\label{inftytransfo}
\Delta T_\mu&=\partial_\mu b-T_\nu\partial_\mu\xi^\nu,\\
\notag
\Delta A_\mu&=\partial_\mu\theta-A_\nu\partial_\mu\xi^\nu+mu_a S^a_\mu-ma_aB^a_\mu,
\end{align}
where the internal vector indices are lowered and raised by the Kronecker metrics $\delta_{ab}$ and $\delta^{ab}$. By construction, all gauge fields transform as one-forms under coordinate diffeomorphisms. For the sake of future reference, we add explicit expressions for the field-strength tensor, defined in the matrix form as $\A_{\mu\nu}\equiv\partial_\mu\A_\nu-\partial_\nu\A_\mu-\imag[\A_\mu,\A_\nu]$. By projecting it to the respective generators, one obtains the field strengths of the individual gauge fields,
\begin{align}
\notag
R^{ab}_{\mu\nu}&=\partial_\mu R^{ab}_\nu-\partial_\nu R^{ab}_\mu+R^{ac}_\mu R^b_{\nu c}-R^{ac}_\nu R^b_{\mu c},\\
\label{fieldstrength}
B^a_{\mu\nu}&=\partial_\mu B^a_\nu-\partial_\nu B^a_\mu-R^a_{\mu b}B^b_\nu+R^a_{\nu b}B^b_\mu,\\
\notag
S^a_{\mu\nu}&=\partial_\mu S^a_\nu-\partial_\nu S^a_\mu-R^a_{\mu b}S^b_\nu+R^a_{\nu b}S^b_\mu-B^a_\mu T_\nu+B^a_\nu T_\mu,\\
\notag
T_{\mu\nu}&=\partial_\mu T_\nu-\partial_\nu T_\mu,\\
\notag
A_{\mu\nu}&=\partial_\mu A_\nu-\partial_\nu A_\mu-mB^a_\mu S_{\nu a}+mB^a_\nu S_{\mu a}.
\end{align}
The transformation rules for the components of the field-strength tensor are obtained from Eq.~\eqref{inftytransfo} by dropping all terms containing a derivative of the parameters $\alpha^{ab}$, $u^a$, $a^a$, $b$, $\theta$.

The physical meaning of some of the gauge fields introduced above is immediately clear, as they represent sources for the respective conserved currents: $S^a_\mu$ for momentum, $T_\mu$ for energy, and $A_\mu$ for the particle number current. In Sec.~\ref{sec:GCI}, we already interpreted fluctuations of the spacetime vielbein as sources for the momentum and energy currents. One of our first tasks therefore will be to compute the MC form to check whether the covariant vielbein obtained from it is compatible with our earlier definition. Furthermore, as we will see, the gauge field $R^{ab}_\mu$ will give us access to spin degrees of freedom analogously to the relativistic case. Finally, the gauge field $B^a_\mu$ in principle couples to the conserved currents for Galilei boosts, although it does not seem to admit a simple physical interpretation.


\subsection{Choice of the coset space}
\label{subsec:cosetchoice}

Before applying the strategy of Sec.~\ref{subsec:poincare} to describe Galilei-invariant systems, it is worth pausing and commenting on the various ways in which a coset construction for the Galilei group can be implemented consistently with unbroken rotational symmetry. While for a relativistic system in vacuum the choice of the coset space $\G/\H$ is essentially unique, the commutation relations of the Galilei algebra~\eqref{commutators} admit several consistent options. Given that spacetime translations are always nonlinearly realized, there are altogether four different options for the choice of generators of $\H$:
\begin{enumerate}
\item ${\{J_{ab},N_a,Q\}}$. This option corresponds to using a coset parametrization identical to the relativistic one given in Eq. (\ref{cosetparam}). While mathematically consistent, it turns out to be impractical: due to the commutation relation $[P_a,N_b]=\imag mQ\delta_{ab}$, the coset generators $\{P_a,H\}$ do not span a representation of the subgroup $\H$. As a consequence, the coset construction cannot be applied directly as the $\omega_{\perp\mu}$ part of the MC form does not transform covariantly like in Eq.~\eqref{MCtransfo}. 
\item ${\{J_{ab},N_a\}}$. Here one chooses the generators of $\mathcal{H}$ as in the Poincar\'e case. The remaining generators $\{P_a,H,Q\}$ form a representation of $\H$ which is reducible but \emph{indecomposable}~\cite{Niederle:galilei}. It is still possible to build manifestly Galilei-covariant expressions using such a representation provided one embeds the Galilei group in the Lorentz group of a spacetime with one extra dimension (see Ref.~\cite{Santos:2004pq} and references therein). Within this formalism, the coset generators $\{P_a,H,Q\}$ define a vector of $\H$ with an extra component due to $Q$. As a result, tensor fields (such as gauge fields) now have more components, some of which are unphysical and have to be eliminated in a covariant manner.
\item ${\{J_{ab},Q\}}$. Thanks to $Q$'s inclusion in the subgroup $\H$, this option allows the straightforward inclusion of charged matter fields. Additionally, it automatically leads to the correct Galilei transformation for such fields, whereby a time-dependent coordinate shift is accompanied by a change of phase of the field. Since Galilei boosts $N_a$ belong to the coset space $\G/\H$, we will have to introduce a set of fields $v^a(x)$ that account for their nonlinear realization. The resulting EFT framework naturally reproduces a version of Schr\"odinger field theory with an auxiliary velocity field which makes the equations of motion of first order in derivatives. This choice will be discussed in detail in Sec.~\ref{sec:microscopic}.
\item ${\{J_{ab}\}}$. In this scheme, internal U(1) transformations generated by $Q$ are realized nonlinearly and consequently one needs to introduce another field, $\pi(x)$. The boost fields $v^a$ are still present, but in this case can be eliminated algebraically by a set of covariant conditions, known as the inverse Higgs constraints~\cite{Ivanov:1975zq,Endlich:2013vfa,Brauner:2014aha}. Since the coset fields can always be interpreted as Nambu--Goldstone (NG) fields of spontaneously broken symmetry, this setup is ready-made for an EFT description of superfluid phases of matter, in which particle number is spontaneously broken. This implementation of the coset construction will be described in Sec.~\ref{sec:superfluid}. 
\end{enumerate}


\section{Microscopic theory}
\label{sec:microscopic}

Here we develop the framework corresponding to a subgroup $\H$ that is generated by $J_{ab}$ and $Q$. After making this choice for the coset space, the mathematical structure of the setup is completely dictated by symmetry. We therefore first work out the details and only then interpret what we found.

Following closely the general discussion in Appendix~\ref{app:coset}, we parametrize the coset space $\G/\H$ as
\begin{equation}
U(x)\equiv e^{\imag y^a(x)P_a}e^{\imag z(x)H}e^{\imag v^a(x)N_a}.
\label{Udef}
\end{equation}
The gauged MC form is defined as usual by Eq.~\eqref{MCformgeneral}. In order to evaluate it explicitly, we make use of the conjugation relations that follow from the commutators~\eqref{commutators},
\begin{equation}
\begin{split}
e^{-\imag\vek\alpha\cdot\vek P}J_{ab}e^{\imag\vek\alpha\cdot\vek P}&=J_{ab}-\alpha_aP_b+\alpha_bP_a,\\
e^{-\imag\vek\alpha\cdot\vek P}N_ae^{\imag\vek\alpha\cdot\vek P}&=N_a + m\alpha_aQ,\\
e^{-\imag\vek\alpha\cdot\vek N}P_ae^{\imag\vek\alpha\cdot\vek N}&=P_a - m\alpha_aQ,\\
e^{-\imag\vek\alpha\cdot\vek N}He^{\imag\vek\alpha\cdot\vek N}&=H - \vek\alpha\cdot\vek P+\tfrac12m\vek\alpha^2Q,
\end{split}
\label{Nconj}
\end{equation}
where we used the shorthand notation $\vek u\cdot\vek v\equiv\delta_{ab}u^av^b$. The individual components of the MC form, projected onto the respective generators, read
\begin{align}
\notag
\omega_\mu^{Jab}=&-R^{ab}_\mu,\\
\label{MCunbroken}
\omega_\mu^Q=&-A_\mu-m\vek y\cdot\vek B_\mu-m\vek v\cdot\partial_\mu\vek y\\
\notag
&+mv_a(S^a_\mu+R^{ab}_\mu y_b+zB^a_\mu)+\tfrac12m\vek v^2(\partial_\mu z-T_\mu),
\end{align}
for the generators of $\H$, and
\begin{align}
\notag
\omega_\mu^{Na}&=\partial_\mu v^a-B^a_\mu-R^{ab}_\mu v_b,\\
\notag
\omega_\mu^{Pa}&=\partial_\mu y^a-(S^a_\mu+R^{ab}_\mu y_b+zB^a_\mu)-v^a(\partial_\mu z-T_\mu),\\
\label{MCbroken}
\omega^H_\mu&=\partial_\mu z-T_\mu.
\end{align}
for the generators of $\G/\H$.


\subsection{Fields and symmetry transformations}
\label{subsec:fieldstransfo}

In order to understand the structure just introduced, it is important to work out the transformation properties of all the ingredients. The canonical transformation of coset fields is given by Eq.~\eqref{cosettransfo2}, whereas that of the MC form by Eq.~\eqref{MCtransfo}. From these general results, it is easy to deduce the following transformation properties.

\textit{Spatial rotations.} These are by construction realized linearly on the coset fields: $z$ is a scalar while $y^a$ and $v^a$ are vectors. In this case, $\h=\g=e^{\frac\imag2\alpha^{ab}J_{ab}}$. The components of the MC form $\omega^Q_\mu$ and $\omega^H_\mu$ are scalars while $\omega^{Na}_\mu$ and $\omega^{Pa}_\mu$ are vectors under rotations. Finally, $\omega^{Jab}_\mu$ transforms as a tensor gauge field, just like $-R^{ab}_\mu$.

\textit{U(1) charge transformations.} Here we have again $\h=\g=e^{\imag\theta Q}$; all coset fields are trivially invariant. Likewise, all components of the MC form remain U(1)-invariant except for $\omega^Q_\mu$ which behaves as a gauge field,
\begin{equation}
\delta_Q\omega^Q_\mu=-\partial_\mu\theta.
\end{equation}

\textit{Space and time translations.} In the parametrization given by Eq.~\eqref{Udef}, the coset fields transform simply,
\begin{equation}
\delta_{P,H}y^a=a^a,\qquad
\delta_{P,H}z=b,\qquad
\delta_{P,H}v^a=0.
\end{equation}
Consequently, $\h=\openone$ and all components of the MC form are trivially invariant under translations.

\textit{Galilei boosts.} This is the most interesting transformation. By a direct computation using Eq.~\eqref{Nconj}, we find that $z$ is boost-invariant whereas
\begin{equation}
\delta_Ny^a=zu^a,\qquad
\delta_Nv^a=u^a.
\end{equation}
As expected, $z$ and $y^a$ behave exactly as time and a spatial coordinate, while $v^a$ behaves as a velocity. Furthermore, we find that $\h=e^{\imag m(\vek u\cdot\vek y+\frac12z\vek u^2)Q}$. Since this matrix defines the transformation rule for matter fields, we recover the well-known fact that for any charged field, a Galilei boost is accompanied by a change of its phase. Due to the form of $\h$, we finally observe that all components of the MC form are boost-invariant except for $\omega^Q_\mu$ which changes under an infinitesimal boost as
\begin{equation}
\delta_N\omega^Q_\mu=-m\partial_\mu(\vek u\cdot\vek y).
\end{equation}


\subsection{Invariant actions}
\label{subsec:actions}

The coset construction provides us with all the ingredients we need to build manifestly invariant actions. As to the covariant vielbein $e^A_\mu$, we naturally identify
\begin{equation}
n_\mu=\omega^H_\mu,\qquad
e^a_\mu=\omega^{Pa}_\mu.
\end{equation}
Given a matter field $\psi$, the $\H$-components of the MC form allow us to define its covariant derivative, 
\begin{equation}
D_\mu\psi=[\partial_\mu+\imag q\omega^Q_\mu-\tfrac{\imag}{2} R_\mu^{ab} \mathcal R(J_{ab})]\psi,
\end{equation}
where $q$ is the electric charge of $\psi$, and $\mathcal R$ the (spin) representation of the rotation group in which it transforms. The dual vielbein $E^\mu_A$ can in turn convert $D_\mu\psi$ into a spacetime scalar, see Eq.~\eqref{scalarderivative}. 

We have managed to identify the building blocks that were used in Secs.~\ref{sec:GCI} and~\ref{sec:NRGCI} to construct the action. However, the coset construction tells us much more. Namely, we now know how to contract the vielbein indices: they have to be summed over in  a way that preserves the linearly realized subgroup $\H$, that is, rotations and U(1) phase redefinitions. Notice that Galilei invariance is already built in automatically, thanks to the presence of the coset field $v^a$.  

For a complex scalar field $\psi$, there is a \emph{unique} Lagrangian that respects all the symmetries and contains just one derivative,
\begin{equation}
\La=\frac\imag2\he\psi\symd\DD_0\psi-V(\he\psi\psi).
\label{Lagaux}
\end{equation}
This should be contrasted with Eq.~\eqref{LagImproved}. In order to see that the two Lagrangians describe equivalent theories, we have to deal with the boost NG field $v^a$ hidden in the definition of $E_A^\mu$ and $\omega^Q_\mu$ . To highlight the precise way this NG field enters the Lagrangian~\eqref{Lagaux}, it is convenient to introduce the ``bare'' vielbein with $v^a$ removed,
\begin{equation}
\bar e^a_\mu\equiv\partial_\mu y^a-(S^a_\mu+R^{ab}_\mu y_b+zB^a_\mu),
\end{equation}
so that $e^a_\mu=\omega^{Pa}_\mu=\bar e^a_\mu-v^an_\mu$. Denoting the ``bare'' dual vielbein, inverse to $\bar e^a_\mu$ and $n_\mu$, analogously as $\bar E^\mu_a,\bar V^\mu$, it is easy to verify the relations
\begin{equation}
E^\mu_a=\bar E^\mu_a,\qquad
V^\mu=\bar V^\mu+v^a\bar E^\mu_a.
\label{EVdual}
\end{equation}
Finally, define the modified gauge field $\bar A_\mu\equiv A_\mu+m\vek y\cdot\vek B_\mu$ so that $-V^\mu\omega^Q_\mu=V^\mu\bar A_\mu+\tfrac12m\vek v^2$. The Lagrangian~\eqref{Lagaux} thus acquires the explicit form
\begin{equation}
\La=\frac\imag2V^\mu\he\psi\symd\partial_{\!\mu}\psi+V^\mu\he\psi\bar A_\mu\psi+\tfrac12m\vek v^2\he\psi\psi-V(\he\psi\psi).
\label{Lagaux2}
\end{equation}
The field $v^a$ is obviously not dynamical because it enters Eq.~\eqref{Lagaux2} without any derivatives. Upon integrating $v^a$ out using its equation of motion~\cite{Hoyos:2013eha}, the Lagrangian~\eqref{LagImproved} is recovered, provided $\bar e^a_\mu$ and $\bar A_\mu$ are identified with the vielbein and the U(1) gauge field introduced therein~\footnote{Apart from the terms present in the Lagrangian~\eqref{LagImproved}, a nonlocal operator proportional to $g^{\mu\nu}\partial_\mu\rho\partial_\nu\rho/\rho$ with $\rho\equiv\psi^\dagger\psi$ appears upon integrating $v^a$ out. As long as the particle number U(1) symmetry is spontaneously broken by a nonzero vacuum expectation value of $\psi$, this operator will be irrelevant at energies well below the symmetry breaking scale. Being separately invariant under all the symmetries of the problem, we can simply discard it without affecting the symmetry properties of the action.}.

It is integrating out the boost NG field $v^a$ that eventually ensures that all the ingredients introduced in Sec.~\ref{sec:NRGCI} have appropriate transformation rules under Galilei boosts and appear in the Lagrangian in the right combinations. While $e^A_\mu$ is boost-invariant, it follows from Eq.~\eqref{EVdual} that $\delta_N\bar V^\mu=-u^a\bar E^\mu_a$. Likewise, it is easy to check the transformation rule for $\bar A_\mu$,
\begin{equation}
\Delta\bar A_\mu=\partial_\mu(\theta+m\vek u\cdot\vek y)-\bar A_\nu\partial_\mu\xi^\nu-mu_a\bar e^a_\mu.
\end{equation}
From here, one immediately recovers the previously guessed rule~\eqref{galileiguess} for covariant derivatives together with the phase factor induced by Galilei boosts as appropriate for the complex field $\psi$.


\subsection{Relations among conserved currents}
\label{subsec:relations}

As we have just shown, the coset formalism exactly reproduces the results of Sec.~\ref{sec:NRGCI}. Moreover, we now have access to a wealth of information due to the presence of independent sources for translation, rotation, boost and electromagnetic currents. Eqs.~\eqref{MCunbroken} and~\eqref{MCbroken} ensure that the sources appear only through certain specific combinations. For instance, $A_\mu$ always appears together with the boost source in the linear combination $\bar A_\mu$, defined above, while $S^a_\mu$ always enters through the combination
\begin{equation}
\bar S^a_\mu\equiv S^a_\mu+R^{ab}_\mu y_b+zB^a_\mu,
\end{equation}
so that $\bar e^a_\mu=\partial_\mu y^a-\bar S^a_\mu$. The combined diffeomorphism--gauge transformation of this field reads
\begin{equation}
\Delta\bar S^a_\mu=\partial_\mu(a^a+\alpha^{ab}y_b+zu^a)-\bar S^a_\nu\partial_\mu\xi^\nu-\alpha^a_{\phantom ab}\bar e^b_\mu-u^an_\mu,
\end{equation}
which indicates that $\bar S^a_\mu$ transforms as a gauge field with respect to the \emph{total} coordinate shift, including the contributions from rotations and boosts.

Let us initially assume that the theory does not contain any fields with spin so that $\omega^J_\mu$ does not contribute, and that $\omega^N_\mu$ can be disregarded (see Sec.~\ref{sec:superfluid} on this point). After integrating out the auxiliary field $v^a$, the action then quite generally depends on the sources only through $n_\mu$, $\bar e^a_\mu$ and $\bar A_\mu$. This eventually leads to the relations
\begin{equation}
M^{\mu ij}=x^iT^{\mu j}-x^jT^{\mu i},\qquad
B^{\mu i}=tT^{\mu i}-mx^ij^\mu
\end{equation}
among the particle number current $j^\mu$, momentum current $T^{\mu i}$, boost current $B^{\mu i}$ and angular momentum current $M^{\mu ij}$, valid in Cartesian coordinates in flat spacetime~\cite{Brauner:2014aha}. These identities directly reflect the fact that a local rotation can be compensated by a local translation, and a local Galilei boost can be compensated by a combination of a local translation and a local U(1) transformation~\cite{Low:2001bw}. Note that the identity~\eqref{T0imji} is \emph{not} a relation between currents, equating momentum density to the U(1) current. It is also a consequence of Galilei symmetry though, and can be understood as a consistency condition associated with the above identity for $B^{\mu i}$~~\cite{Brauner:2014aha}.


\subsection{Spin connection and torsion}
\label{subsec:spin}

So far, we have simplified the discussion by considering solely fields without spin. However, the coset construction provides a concise description of the general case. To that end, it is suitable to trade the components of the MC form for their gauge-covariant combinations, given by the field-strength tensor, $\omega_{\mu\nu}\equiv\partial_\mu\omega_\nu-\partial_\nu\omega_\mu+\imag[\omega_\mu,\omega_\nu]$. After some manipulations using Eqs.~\eqref{fieldstrength}, \eqref{MCunbroken} and~\eqref{MCbroken}, its components can be given the explicit form,
\begin{align}
\notag
\omega^{Jab}_{\mu\nu}=&-R^{ab}_{\mu\nu},\\
\notag
\omega^Q_{\mu\nu}=&-A_{\mu\nu}-my_aB^a_{\mu\nu}+mv_a(S^a_{\mu\nu}+R^{ab}_{\mu\nu}y_b+zB^a_{\mu\nu})\\
&-\tfrac12m\vek v^2T_{\mu\nu},\\
\notag
\omega^{Na}_{\mu\nu}=&-B^a_{\mu\nu}-R^{ab}_{\mu\nu}v_b,\\
\notag
\omega^{Pa}_{\mu\nu}=&-S^a_{\mu\nu}-R^{ab}_{\mu\nu}y_b-zB^a_{\mu\nu}+v^aT_{\mu\nu},\\
\notag
\omega^H_{\mu\nu}=&-T_{\mu\nu}.
\end{align}
In analogy with the relativistic case, we can therefore interpret $\omega^{Jab}_{\mu\nu}$ as the \emph{spatial curvature} tensor and $\omega^H_{\mu\nu}$ as the \emph{temporal torsion} tensor (both up to a sign).

The other components of the MC form depend on the dynamical field $v^a$ so we have to be more careful. We define the \emph{spatial torsion} tensor by stripping off the $v^a$-term from $\omega^{Pa}_{\mu\nu}$,
\begin{equation}
\mathscr S^a_{\mu\nu}\equiv S^a_{\mu\nu}+R^{ab}_{\mu\nu}y_b+zB^a_{\mu\nu}.
\label{spatialtorsion}
\end{equation}
It is easy to check using Eq.~\eqref{inftytransfo} that $\mathscr S^a_{\mu\nu}$ is covariant under internal rotations and invariant under internal spacetime translations and U(1) transformations. The only transformation that affects it is the internal Galilei boost, under which $\delta_N\mathscr S^a_{\mu\nu}=u^aT_{\mu\nu}$. As could have been expected, $\mathscr S^a_{\mu\nu}$ together with $T_{\mu\nu}$ transform as a vector under Galilei boosts.

To further check the consistency of the definition~\eqref{spatialtorsion}, we note that $\mathscr S^a_{\mu\nu}$ can be expressed in terms of the MC form rather than in terms of the field-strength tensor $\mathcal A_{\mu\nu}$ in a way independent of $v^a$,
\begin{equation}
\mathscr S^a_{\mu\nu}=-\partial_\mu\bar e^a_\nu+\partial_\nu\bar e^a_\mu+R^a_{\mu b}\bar e^b_\nu-R^a_{\nu b}\bar e^b_\mu+n_\mu B^a_\nu-n_\nu B^a_\mu.
\label{eliminate}
\end{equation}
Barring the appearance of the boost source $B^a_\mu$, the generating functional of the EFT depends on the vielbein $n_\mu$, $\bar e^a_\mu$, electromagnetic source $\bar A_\mu$, and the spin connection $R^{ab}_\mu$. Owing to the fact that Eq.~\eqref{eliminate} is algebraic in $R^{ab}_\mu$, the latter can be traded for the spatial torsion $\mathscr S^a_{\mu\nu}$~\cite{Bradlyn:2014wla}. Since the spatial torsion depends on the choice of reference frame, covariant constraints on the background can be obtained by setting either the temporal torsion, or both the temporal and the spatial torsion to zero. In the latter case, the spin connection can be expressed in terms of derivatives of the vielbein as in Eq.~\eqref{spinconnection}.


\subsection{Rotationally invariant systems}

The formalism developed above allows a streamlined construction of effective actions invariant under internal U(1) symmetry, spacetime translations, spatial rotations and Galilei boosts. In real condensed matter systems, boost invariance is often broken at a much higher energy scale than other symmetries. It is therefore illustrative to inspect how our results modify in this case, where the low-energy EFT is invariant under spatial rotations but not under the boosts.

Relaxing the constraints imposed by boost invariance is, in fact, extremely simple. In all the above formulas we have to discard the external source $B^a_\mu$ as well as the coset field $v^a$. The MC form thereby reduces to
\begin{equation}
\begin{gathered}
\omega^{Jab}_\mu=-R^{ab}_\mu,\qquad
\omega^Q_\mu=-A_\mu,\\
\omega^{Pa}_\mu=\partial_\mu y^a-\bar S^a_\mu,\qquad
\omega^H_\mu=\partial_\mu z-T_\mu,
\end{gathered}
\end{equation}
where $\bar S^a_\mu=S^a_\mu+R^{ab}_\mu y_b$. The temporal torsion $T_{\mu\nu}$ and spatial torsion $\mathscr S^a_{\mu\nu}$ become completely decoupled, depending just on $n_\mu$ and on $\bar e^a_\mu$ and $R^{ab}_\mu$, respectively. Eq.~\eqref{eliminate} becomes identical to a Euclidean version of its relativistic counterpart. If desired, it can be used to eliminate the spin connection in favor of the vielbein as usual. Local invariant Lagrangians are constructed as before by imposing the linearly realized symmetry $\H$, consisting of spatial rotations and internal U(1) transformations.


\section{Superfluid effective theory}
\label{sec:superfluid}

In the previous section, we showed how to construct invariant actions for charged matter fields. The setup therefore provides a suitable tool for the discussion of symmetries of microscopic theories of electrons and nuclei, atoms or molecules. Consider now a system which becomes superfluid at sufficiently low temperatures. Provided there are no other gapless modes in the spectrum, we then expect the low-energy physics to be dominated by the ensuing NG boson. In this case, both Galilei boosts and the particle number U(1) \emph{are} spontaneously broken by the physical ground state, and the framework based on the isotropy subgroup $\H$ generated by $J_{ab}$ becomes appropriate. In this section, we describe the main differences compared to the coset construction carried out in the previous section. The coset construction for a nonrelativistic superfluid in flat spacetime has also been discussed recently in Ref.~\cite{Endlich:2013spa}.

On account of having an extra broken generator, the associated NG field, $\pi(x)$, must be added to the coset element~\eqref{Udef},
\begin{equation}
U(x)\equiv e^{\imag y^a(x)P_a}e^{\imag z(x)H}e^{\imag v^a(x)N_a}e^{\imag\pi(x)Q}.
\end{equation}
Since $Q$ commutes with all other generators, this modification has a limited impact on the results derived in Sec.~\ref{sec:microscopic}. The only change to the MC form is an extra term $\partial_\mu\pi$ in $\omega^Q_\mu$, so that now [cf.~Eq.~\eqref{MCunbroken}]
\begin{equation}
\tilde\omega^Q_\mu=\partial_\mu\pi-\bar A_\mu-m\vek v\cdot\vek e_\mu-\tfrac12m\vek v^2n_\mu.
\end{equation}
The transformation rules of the fields change accordingly. First, under U(1) charge transformations, $\delta_Q\pi=\theta$ so that $\h=\openone$ and consequently all components of the MC form are trivially invariant under U(1). Second, under small Galilei boosts, one now has $\delta_N\pi=m\vek u\cdot\vek y$. Again, $\h=\openone$ and the MC form is completely boost-invariant.

The transformation rules have thus become extremely simple. Spacetime translations, U(1) transformations and Galilei boosts are all realized by shifts of their respective coset fields (plus the  corrections $\delta_N\pi$ and $\delta_Ny^a$, induced by the boosts), leaving the MC form invariant. Spatial rotations, on the other hand, act linearly on all the fields except for the gauge field $R^{ab}_\mu$, as expected.

The EFT now contains two dynamical NG fields: $\pi$ and $v^a$. However, it is well-known that there are no physical gapless states in superfluids that could be associated with the spontaneously broken Galilei boosts. The low-energy spectrum only contains one gapless state, to which \emph{both} the U(1) current and the boost currents couple~\cite{Watanabe:2013iia}. In the EFT, this is reflected by the fact that the $v^a$ field is not protected by symmetry from acquiring a mass term~\cite{Brauner:2014aha,Endlich:2013vfa}. It therefore does not contribute to low-energy physics even if it was initially introduced as an independent degree of freedom.

To obtain an EFT for the physical mode $\pi$ (in the presence of background fields) alone, one eliminates $v^a$ by integrating it out, or by imposing an algebraic ``inverse Higgs'' constraint~\cite{Ivanov:1975zq}. The latter is more convenient as the elimination is performed on the level of the covariant constituents of the theory, without the need to know the details of its action. While the precise form of the inverse Higgs constraint may sometimes be ambiguous, in this case a convenient choice is $E^\mu_a\tilde\omega^Q_\mu=0$, leading to
\begin{equation}
v_a=\frac1mE^\mu_aD_\mu\pi,
\end{equation}
where we set $D_\mu\pi\equiv\partial_\mu\pi-\bar A_\mu$. Plugging this back into the expression~\eqref{EVdual} for the dual vielbein, it becomes
\begin{equation}
E^\mu_a=\bar E^\mu_a,\qquad
V^\mu=\bar V^\mu+\frac{1}{m}g^{\mu\nu}D_\nu\pi,
\end{equation}
where we used the degenerate metric $g^{\mu\nu}$, introduced in Eq.~\eqref{gmunu}. Since the low-energy spectrum contains no gapped matter states on which $\omega^J_\mu$ could act, the only building blocks at our disposal are
\begin{equation}
\begin{split}
\omega^{Na}_\mu&=-B^a_\mu+\frac1mD_\mu(E^{a\nu}D_\nu\pi),\\
V^\mu\tilde\omega^Q_\mu&=\bar V^\mu D_\mu\pi+\frac{g^{\mu\nu}}{2m}D_\mu\pi D_\nu\pi,
\end{split}
\end{equation}
where, in the last term on the right-hand side of the first line, $D_\mu\psi^a\equiv\partial_\mu\psi^a-R^a_{\mu b}\psi^b$.

In a power-counting scheme where $\partial_\mu\pi$ counts as order zero and any additional derivative acting on the fields increases the order~\cite{Son:2002zn}, $\omega^{N}_\mu$ will not contribute at the lowest order in derivatives. The leading-order action will then be given by~\footnote{The $g$ in the volume measure here of course refers to the spatial metric $g_{ij}$, defined by Eq.~\eqref{gijdef}. The opposite sign in front of the $\frac{g^{\mu\nu}}{2m}D_\mu\pi D_\nu\pi$ term as compared to Ref.~\cite{Son:2005rv} is just a matter of a sign convention for the field $\pi$.}
\begin{equation}
S=\int\dd t\,\dd\vek x\,n_0\sqrt{\|g\|}\,P(V^\mu\tilde\omega^Q_\mu),
\end{equation}
where $P$ defines the thermodynamic pressure of the system at zero temperature as a function of the chemical potential for the U(1) symmetry. This generalizes the action found in Ref.~\cite{Son:2005rv} to an arbitrary spacetime background, including sources for momentum density and energy density and current.


\section{Conclusions}
\label{sec:conclusions}

In this paper, we have developed a framework that casts a given Lagrangian field theory in a form manifestly invariant under arbitrary coordinate reparametrizations, and at the same time provides a transparent treatment of symmetries of the system. By making systematic use of the vielbein formalism, we naturally reproduced in Secs.~\ref{sec:GCI} and~\ref{sec:NRGCI} the Newton--Cartan structure, recently introduced in the studies of the fractional quantum Hall effect. (See also Ref.~\cite{Christensen:2013lma,*Christensen:2013rfa} for a study of Newton--Cartan geometry with torsion in a different context.) Utilizing the coset construction, we then showed in Secs.~\ref{sec:galilei},~\ref{sec:microscopic} and~\ref{sec:superfluid} how the physical symmetries can be implemented separately from coordinate diffeomorphisms. The discussion in Sec.~\ref{subsec:actions} suggests that it is more natural to construct Galilei-invariant actions with the additional auxiliary velocity field $v^a$. Upon integrating this out, the action takes a form that seems difficult to obtain directly in a systematic manner.

Although our approach might seem intimidating due to the amount of notation and algebra involved, it is in fact extremely simple conceptually: once the basic scheme is set up, all the algebraic manipulations are enforced by symmetry, and do not require any further insight or guesswork. To conclude, we would like to make several comments on the potential extensions and applications of our approach in the form of final remarks.

First, we used for illustrative purposes only the simplest type of model: a NR Galilei-invariant system with a U(1) internal symmetry. This U(1) symmetry plays a special role in that it enters the commutation relations of the Galilei algebra~\eqref{commutators}. As a consequence, the associated gauge field $A_\mu$ enters the action in a nontrivial manner. We anticipate that other, possibly non-Abelian, internal symmetries can be added straightforwardly, as suggested in Ref.~\cite{Andreev:2013qsa}. This is obvious in the coset formalism, worked out in Sec.~\ref{sec:microscopic}: adding an extra set of generators that commute with the Galilean algebra results in a separate contribution to the MC form which does not interfere with already existing structure.

Second, the formalism can straightforwardly be applied to any local field theory regardless of its symmetries and particle composition. This underlines the fact that there is very little physical content in general coordinate invariance alone, expressing merely the freedom to choose an arbitrary coordinate system for the description of physical observables. An interesting direction of future work would be to apply the formalism developed here to systems where spacetime symmetries such as rotations or translations actually are spontaneously broken. This would represent a synthesis of the approach of Refs.~\cite{Nicolis:2013lma,Delacretaz:2014jka}, where EFTs for NG modes of spacetime symmetries were studied, with that of Refs.~\cite{CSpaper,EFTpaper} which applied the generating functional technique to spontaneously broken internal symmetries.

Third, note that in two spatial dimensions, the Galilei algebra possesses another, exotic central charge~\cite{Jackiw:2000tz,*Jackiw:2002he,*Hagen:2002pg}, possibly related to the two-dimensional spin. While the inclusion of such a central charge by means of an additional U(1) generator in the coset construction seems straightforward, it would be interesting to investigate its physical consequences for, say, two-dimensional superfluids. 


\section*{Acknowledgments}

We acknowledge correspondence and conversation with Kristan Jensen, which prompted us to think of our results from a different perspective. T.B.~is indebted to Sergej Moroz for illuminating discussions. He also gratefully appreciates the hospitality of EPFL, where this project was initiated. S.E.~and R.P.~would like to thank Rachel Rosen for stimulating discussions and for collaboration during the early stages of this project. T.B., S.E.~and R.P.~furthermore acknowledge the hospitality of the Perimeter Institute for Theoretical Physics during the workshop ``Low Energy Challenges for High Energy Physicists'', which gave considerable momentum to our collaboration. The work of T.B.~was supported by the Austrian Science Fund (FWF), Grant No.~M 1603-N27. The work of A.M.~was supported by the Swiss National Science Foundation. The work of R.P.~was supported by NASA under contract NNX10AH14G and by the DOE under contract DE-FG02-11ER41743. During the final stages of this work S.E.~was a Sitka Fellow in Sitka, AK, U.S.A., and gratefully acknowledges the support of the Island Institute.


\appendix
\section{Overview of the coset construction}
\label{app:coset}

Suppose that the action of a given theory is invariant under the Lie group $\G$. It is customary to assume that elementary fields that enter the action span multiplets of this group, that is, the symmetry transformations from $\G$ act on them \emph{linearly}. However, this is not always the case. For instance, spacetime translations act on the coordinates $x^\mu$ by a shift, $x'^\mu=x^\mu+\xi^\mu$. Likewise, NG fields of a spontaneously broken symmetry transform by a similar shift under the broken transformations, which in turn guarantees the low-energy theorems for the associated NG bosons~\cite{Burgess:1998ku}. As soon as a symmetry is realized nonlinearly, invariant actions cannot be obtained by usual tensor methods, that is, by taking a product of fields and then contracting their indices with invariant tensors of the symmetry group. The coset construction~\cite{Coleman:1969sm,*Callan:1969sn} solves the problem, and we outline here its essentials needed in the body of the paper.

We assume that the group $\G$ corresponds to an ``internal'' symmetry in the sense that it does not affect the spacetime coordinates $x^\mu$, and its action on the fields does not depend on $x^\mu$ explicitly. This is reasonable for, as stressed repeatedly throughout the paper, true symmetries can be implemented in a way that does not refer to coordinate reparametrizations, the latter merely representing the freedom to choose a coordinate system. We further need to know the subgroup $\H\subset\G$ which is realized linearly on all the fields.

To find a nonlinear realization of the full group $\G$, one first defines its action on the coset space $\G/\H$. This space consists of all (mutually disjoint) sets of the form $\chi_\g\equiv\{\g\h\,|\,\h\in\H\}$ with fixed $\g\in\G$~\footnote{Since the relation $\g\sim\g'$ indicating the existence of $\h\in\H$ such that $\g'=\g\h$ is an equivalence, any two cosets $\chi_\g$ and $\chi_{\g'}$ are either disjoint or identical. The coset space $\G/\H$ provides a decomposition of the group $\G$ into equivalence classes of $\sim$.}. For each such coset $\chi$, one can pick a fixed representative, $U_\chi$. In other words, every element $\g\in\G$ can be uniquely decomposed as
\begin{equation}
\g=U_{\chi_\g}\h_\g,
\end{equation}
where $\h_\g\in\H$. The action of the group on the coset space is then defined by left multiplication
\begin{equation}
U_{\chi_{\g'}}\xrightarrow{\g}U_{\chi_{\g\g'}}=\g U_{\chi_{\g'}}\h_{\g'}\h^{-1}_{\g\g'}.
\label{cosettransfo1}
\end{equation}
Denote now the generators of $\H$ temporarily as $T_\alpha$ and the generators of $\G/\H$ as $T_a$. At least in some neighborhood of unity, one can represent a given coset as
\begin{equation}
U(\pi)\equiv e^{\imag\pi^aT_a},
\label{cosetparamgeneral}
\end{equation}
where the objects $\pi^a$ serve as local coordinates on the coset space $\G/\H$. The transformation rule~\eqref{cosettransfo1} can then be given the more familiar form,
\begin{equation}
U(\pi)\xrightarrow\g U(\pi')=\g U(\pi)\h(\pi,\g)^{-1}.
\label{cosettransfo2}
\end{equation}
To implement this construction in a quantum field theory, the coset coordinates are interpreted as fields, $\pi^a(x)$. In case of spontaneous symmetry breaking, the subgroup $\H$ corresponds to the symmetry of the physical vacuum and $\pi^a$ represent the ensuing NG bosons. However, the construction is more general and also applies to nonlinear realizations of symmetries such as translations which are not necessarily spontaneously broken. The geometric fields $\pi^a$ then rather play the role of arbitrary but fixed functions that specify the local coordinate frame.

Once the coset structure is made local, it is natural to promote the action of the symmetry group $\G$ to a local one as well, that is, to make $\g$ coordinate-dependent. To achieve manifest covariance under such gauge symmetry, a set of background gauge fields is needed, which can be put together in the one-form matrix-valued variable
\begin{equation}
\A_\mu=\A_\mu^\alpha T_\alpha+\A^a_\mu T_a.
\end{equation}
The group $\G$ is assumed to act upon it as usual in non-Abelian gauge theory,
\begin{equation}
\A_\mu\xrightarrow\g\g\A_\mu\g^{-1}+\imag\g\partial_\mu\g^{-1}.
\label{gaugetransfogeneral}
\end{equation}
The basic building block of the coset construction is the MC form, whose gauge-covariant version reads
\begin{equation}
\omega_\mu\equiv-\imag U^{-1}(\partial_\mu-\imag\A_\mu)U.
\label{MCformgeneral}
\end{equation}
It decomposes as
\begin{equation}
\omega_\mu=\omega_{\parallel\mu}+\omega_{\perp\mu}\equiv\omega^\alpha_{\parallel\mu}T_\alpha+\omega^a_{\perp\mu}T_a.
\end{equation}
Using Eqs.~\eqref{cosettransfo2} and~\eqref{gaugetransfogeneral}, and \emph{assuming} that the generators $T_a$ span a representation of $\H$~\footnote{This can always be achieved for compact Lie algebras by a suitable choice of basis of generators.}, it is easy to verify the transformation rules of the MC form,
\begin{equation}
\omega_{\parallel\mu}\xrightarrow\g\h\omega_{\parallel\mu}\h^{-1}-\imag\h\partial_\mu\h^{-1},\qquad
\omega_{\perp\mu}\xrightarrow\g\h\omega_{\perp\mu}\h^{-1},
\label{MCtransfo}
\end{equation}
where $\h$ is defined by Eq.~\eqref{cosettransfo2}.

So far we have only discussed the fields $\pi^a$ whose presence is enforced by the geometry of the coset space. Nongeometric, or matter, fields can be added to the construction at will though. We will denote such fields collectively as $\psi(x)$. By assumption, they transform under $\H$ in some linear (not necessarily irreducible) representation $\mathcal R$: $\psi\xrightarrow\h\mathcal R(\h)\psi$. This prescription can be promoted to a nonlinear realization of the whole group $\G$ by using the matrix $\h$ from Eq.~\eqref{cosettransfo2}: $\psi\xrightarrow\g\mathcal R(\h(\pi,\g))\psi$. Upon such a symmetry transformation, $\psi$ acquires additional coordinate dependence through the $\pi$-dependence of $\h$ even if $\g$ itself is global. A covariant derivative of $\psi$ is constructed with the help of $\omega_{\parallel\mu}$, which transforms as a gauge connection of $\H$,
\begin{equation}
D_\mu\psi\equiv[\partial_\mu+\imag\mathcal R(\omega_{\parallel\mu})]\psi.
\label{MCformcovder}
\end{equation}

Invariant Lagrangians can be assembled using standard tensor methods out of $\psi$, $\omega_{\perp\mu}$ and their covariant derivatives. In fact, even Lagrangians invariant only up to a surface term, leading to topological actions of the Wess--Zumino type~\cite{DHoker:1994ti,Goon:2012dy,Delacretaz:2014jka,CSpaper}, can be obtained from the building blocks, provided by the MC form. Note that $\omega_{\perp\mu}$ can play a dual role: (i) as a covariant derivative of $\pi^a$ in case these represent dynamical NG fields of a spontaneously broken symmetry; (ii) as a covariant vielbein in case the functions $\pi^a(x)$ represent local coordinates that are not dynamical and can be chosen at will.


\bibliography{references}

\end{document}